\documentclass[aps,prl,showpacs,amsmath,amssymb,twocolumn,floatfix,footinbib,superscriptaddress]{revtex4-1}

\usepackage{amsmath,amssymb}
\usepackage{graphicx}
\usepackage{bm}
\usepackage{bbm}
\usepackage{color}
\usepackage{natbib}
\usepackage{hyperref}
\hypersetup{
  colorlinks,
  citecolor=magenta,
  linkcolor=blue,
  urlcolor=blue}

\newcommand{\cfig}[1]{Fig.~\ref{#1}}

\newcommand{\cref}[1]{Ref.~(\onlinecite{#1})}

\makeatletter
\def\Dated@name{}
\makeatother

\begin{document}
\title{Comment on ``The role of electron-electron interactions in two-dimensional Dirac fermions''}

\author{Stephan Hesselmann}
\affiliation{Institut f\"ur Theoretische Festk\"orperphysik, JARA-FIT and JARA-HPC, RWTH Aachen University, 52056 Aachen, Germany}
\author{Thomas C. Lang}
\affiliation{Institut f\"ur Theoretische Physik, Universit\"at Innsbruck, A-6020 Innsbruck, Austria}
\author{Michael Schuler}
\affiliation{Institut f\"ur Theoretische Physik, Universit\"at Innsbruck, A-6020 Innsbruck, Austria}
\affiliation{Vienna Center for Quantum Science and Technology, Atominstitut, TU Wien, 1040 Wien, Austria}
\author{Stefan Wessel}
\affiliation{Institut f\"ur Theoretische Festk\"orperphysik, JARA-FIT and JARA-HPC, RWTH Aachen University, 52056 Aachen, Germany}
\author{Andreas M. L\"auchli}
\affiliation{Institut f\"ur Theoretische Physik, Universit\"at Innsbruck, A-6020 Innsbruck, Austria}

\begin{abstract}
Tang \textit{et al.} [\href{https://science.sciencemag.org/content/361/6402/570}{Science \textbf{361}, 570 (2018)}] report on the properties of Dirac fermions with both on-site and Coulomb interactions. The substantial decrease up to ${\sim 40\%}$ of the Fermi velocity of  Dirac fermions with on-site interaction is inconsistent with the numerical data near the Gross-Neveu quantum critical point. This results from an inappropriate finite-size extrapolation.
\end{abstract}
\date{Submitted: October 24, 2018, Accepted: November 4, 2019}

\maketitle

The low-energy excitations of many condensed matter systems, such as electrons on the honey-comb lattice of graphene, can be described by massless Dirac fermions with a Dirac cone-like dispersion relation and a corresponding Fermi velocity. The inclusion of interactions among the fermions eventually leads to a breakdown of this description, once the system undergoes a quantum phase transition to an insulating phase beyond a critical interaction strength. Below this interaction-induced quantum critical point (QCP) the system is characterized by massless Dirac fermions with a renormalized Fermi velocity. The quantification of this velocity renormalization constitutes a challenge in numerical simulations: Crossover effects strongly alter finite-size system estimates close to critical points and a careful analysis of the actual excitation energies is required to extract reliable results.

Tang \textit{et al.} \cite{Tang18} extract the momentum resolved one-particle excitation energies from imaginary-time correlation functions obtained by projective quantum Monte Carlo (QMC) simulations. Upon approaching the Dirac points, the lattice dispersion of the non-interacting (tight-binding) fermion system takes on a linear, relativistic form that defines the tight-binding Fermi velocity $v_0$ at the Dirac point. The inclusion of either on-site (Hubbard) interactions or extended Coulomb interactions leads to changes of these excitation energies. Below the interaction-induced Gross-Neveu QCP, the dispersion remains gapless at the Dirac point in the thermodynamic limit (TDL) at infinite lattice size, defining the semi metallic (SM) regime. For the case of the Hubbard model the Gross-Neveu QCP is known to be located at an on-site repulsion of ${U_c(\gamma = 0) = 3.85(2)t}$, beyond which the model exhibits antiferromagnetic order \cite{Otsuka16}. Here, $t$ denotes the nearest neighbor hopping strength on the honeycomb lattice, and ${\gamma = 3\alpha_0/U}$ in terms of the Coulomb interaction strength ${\alpha_0}$. Throughout this comment, we follow the notation used in \cref{Tang18}.

\begin{figure}[t]
  \centering
  \includegraphics[width=\columnwidth]{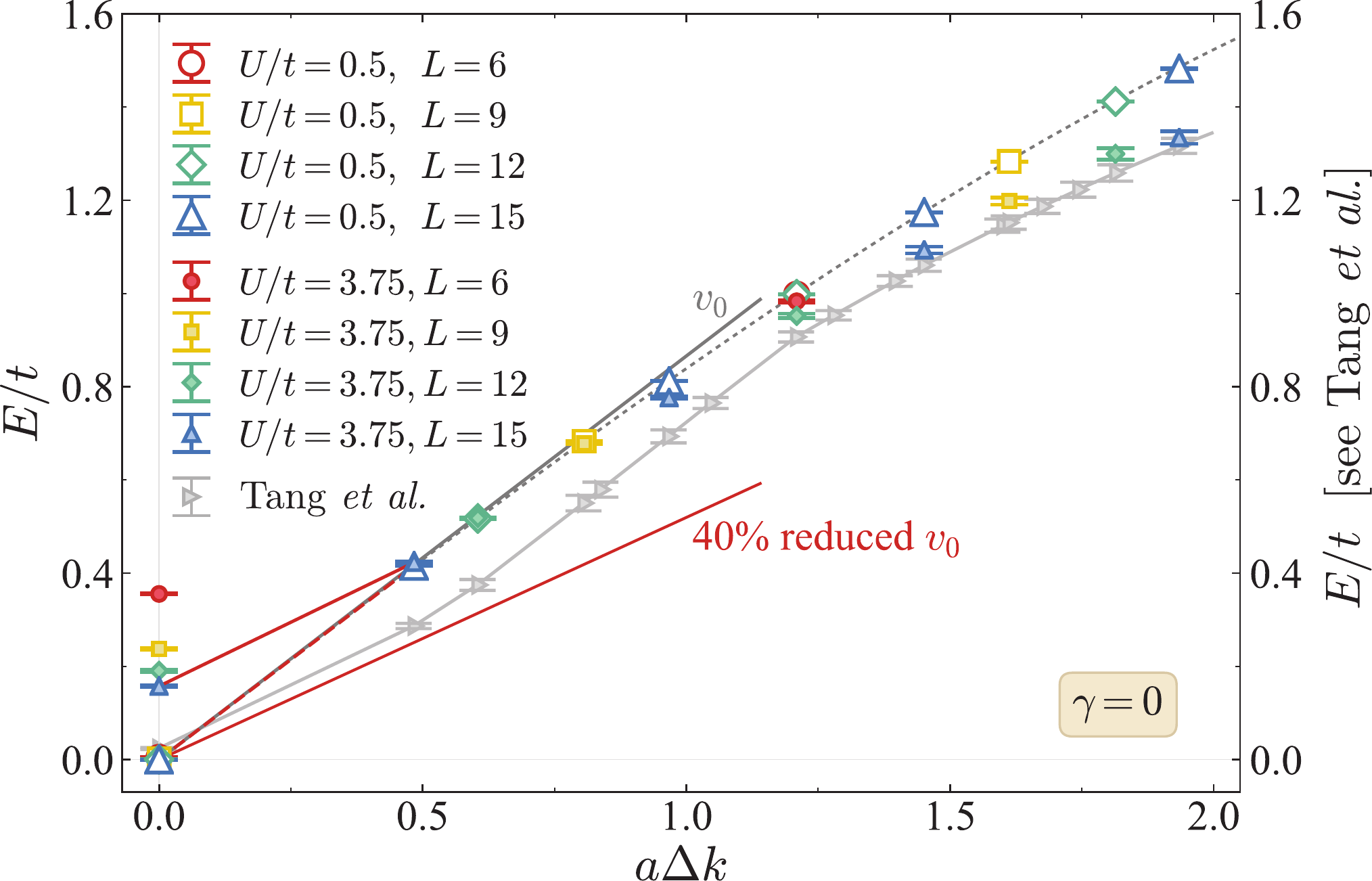}\\
  \caption{Low-energy dispersions for the Hubbard model on the honeycomb lattice at different inter-action strengths. Dependence of the bare lowest particle-excitation energy $E$ on the distance ${a\Delta k}$ to the Dirac point for the Hubbard model (${\gamma = 0}$) on the honeycomb lattice at ${U/t = 0.5}$, and ${U/t = 3.75}$. $E$ is deduced from the imaginary-time slope of the Green's function at the corresponding momenta for different linear lattice sizes $L$ of the system. The dashed dark gray line traces the lattice dispersion relation for the tight-binding model of non-interacting fermions (${U/t = 0}$). Also indicated are linear dispersions corresponding to $v_0$ (dark gray solid line) and to the 40\% decrease with respect to $v_0$ reported in \cref{Tang18} (lower red solid line), and lines that connect the excitation energy at the Dirac point to its value at the nearest neighbor momenta on the ${L = 15}$ lattice for ${U/t = 0.5}$ (dashed red line), and for ${U/t = 3.75}$ (upper solid red line). We include data processed by Tang \textit{et al.} (gray symbols, right scale), which shows their finite size extrapolated gaps for ${U/t=3.75}$ based on the interpolation scheme proposed in their Fig.~S2 of \cref{Tang18} (Supplementary Materials).
  \label{fig1}}
\end{figure}

\begin{figure*}[t]
  \centering
  \includegraphics[width=0.75\textwidth]{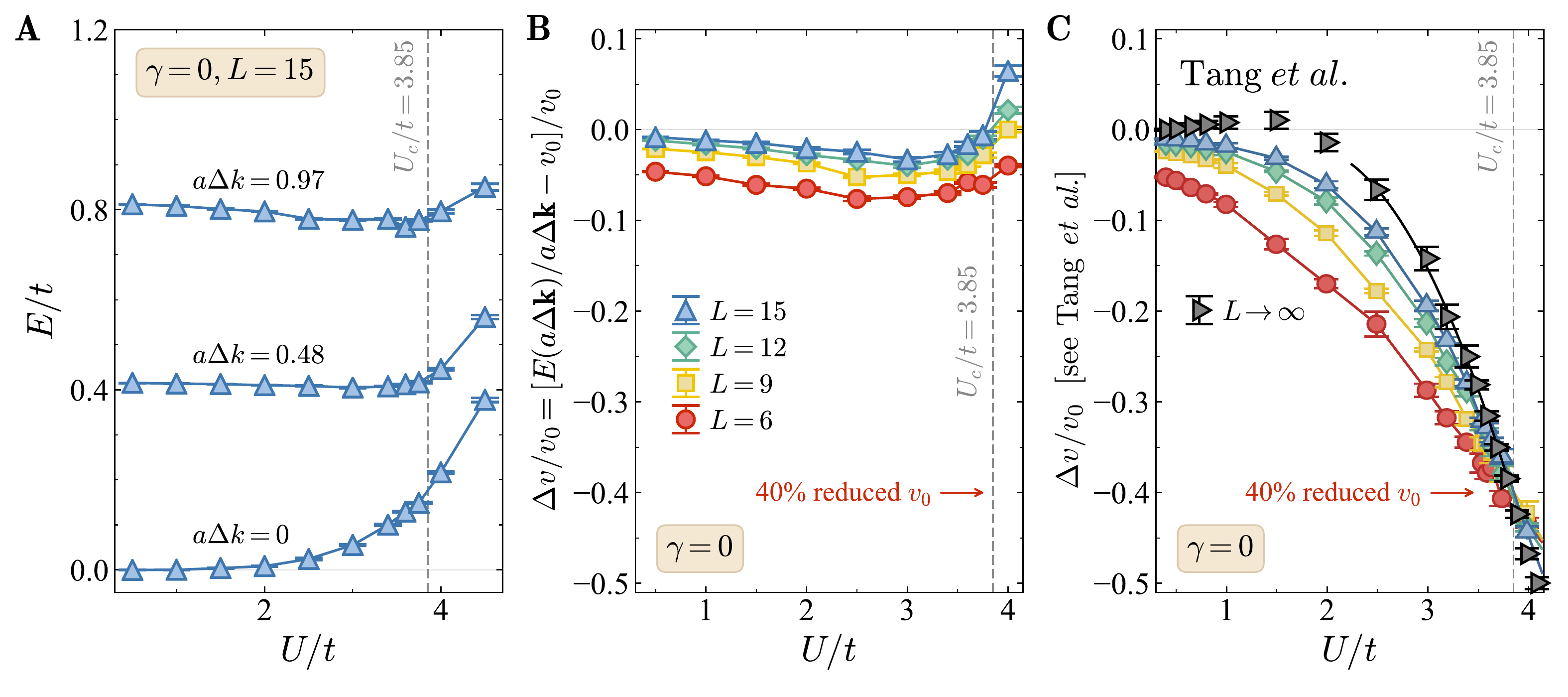}\\
  \caption{Interaction effects on the low-energy excitations for the Hubbard model on the honeycomb lattice.  (A) Dependence of the bare lowest particle-excitation energy E on the strength of the Hub-bard interaction U at the Dirac point ${a\Delta k = 0}$ and at two different distances ${a\Delta = 0.48}$ and 0.97 to the Dirac point for the largest accessed linear system size ${L = 15}$ of \cref{Tang18}. (B) Relative difference between $v_0$ and the rescaled lowest particle-excitation energy ${E(a\Delta k)}$ at the closest momentum to the Dirac point on each finite lattice, as a function of the strength of the Hubbard interaction U for different system sizes L. The red arrow indicates the 40\% decrease with respect to $v_0$ reported in \cref{Tang18}. In both panels the dashed vertical line gives the position of the Gross-Neveu quantum critical point from \cref{Otsuka16}. (C) The estimate for the renormalization of the Fermi velocity as provided by Tang \textit{et al.}, which includes the strongly finite size affected Dirac point. 
  \label{fig2}}
\end{figure*}

In order to extract the interaction-induced renormalization of the Fermi velocity within the SM phase, the excitation gaps obtained from the QMC data for finite-size systems need to be extrapolated to the TDL. Finite-size effects are observed in all excitation energies, but in particular close to the QCP. This is seen in \cfig{fig1}, which shows the bare finite-size excitation gaps, extracted from the imaginary-time QMC data as detailed in the supplementary materials for \cref{Tang18}, based on the data sets made available online by the authors of \cref{Tang18}. We observe that the finite-size effects are most pronounced at the Dirac points themselves (see \cfig{fig1}), where the gap vanishes in the TDL within the SM regime for ${U < U_c(0)}$ and at the Gross-Neveu QCP ${U = U_c(0)}$. On the other hand, for momenta in the immediate vicinity of the Dirac points, the finite-size effects are seen to be much weaker (\cfig{fig1}), and one may estimate the TDL values of the excitation energies at these momenta from the values on the largest system sizes accessed in \cref{Tang18}.

In \cfig{fig1} we also include data provided by Tang \textit{et al.}, showing their finite-size extrapolated gaps. This processed data [based on the interpolation scheme used in their Fig.~S2 (Supplementary Materials)] are seen to be incompatible with the behavior of the excitation energies for small values of ${a\Delta k}$ extracted with our scheme. Moreover, as shown in \cfig{fig2}A, the excitation energies close to, but excluding, the Dirac point exhibit only a weak U-dependence. Thus, for ${\gamma = 0}$, the low-energy Dirac dispersion, and hence the Fermi velocity, is in fact only weakly modified by the on-site inter-actions. In particular, the low-energy dispersion traced by our data in \cfig{fig1} for ${U = 3.75t}$ is clearly inconsistent with the ${\sim 40\%}$ decrease of the Fermi velocity from $v_0$ reported in \cref{Tang18}, which is indicated by the lower red line in \cfig{fig1}. 

A reliable estimate for the Fermi velocity at the Dirac point for values of $U$ inside the SM regime can be obtained from a finite-size analysis of the rescaled lowest particle-excitation energy ${E/(a\Delta k)}$ at the closest momentum to the Dirac point on each finite lattice. The corresponding finite-size values are compared to $v_0$ in \cfig{fig2}B, and demonstrate a remarkably weak renormalization of the Fermi velocity throughout the SM phase. A reduction by ${\sim 40\%}$ from the value $v_0$ is not compatible with the observed steady approach of ${E/(a\Delta k)}$ towards $v_0$ with increasing system size for all considered values of U within the SM regime. 

The substantial overestimation of the Fermi velocity suppression by the on-site interaction reported in \cref{Tang18} (also, see \cfig{fig2}C) is in fact due to an inappropriate finite-size extrapolation procedure, which is documented in Fig.~S2 of \cref{Tang18}: The authors of \cref{Tang18} use the slope between the fi-nite-size excitation energies at the Dirac point and the closest point to the Dirac point (with a linear interpolation to the simulation scale) as estimator. The finite-size energies at the Dirac point suffer from particularly large finite-size effects near the Gross-Neveu QCP, and the strong suppression of the Fermi velocity that is reported in \cref{Tang18} near the Gross-Neveu QCP merely reflects the enhanced finite-size effects of the excitation energy at the Dirac point, but not the renormalization of the actual low-energy dispersion. The extraction of velocities based on the softest excitations is also reported to be subtle for related quantum phase transitions, see, e.g., \cref{Sen15,Schuler16,Schuler19}.  

Their means of data analysis therefore did not allow the authors of \cref{Tang18} to faithfully reproduce the Fermi velocity renormalization beyond the weak-coupling regime. The Fermi velocity renormalization shown in \cfig{fig2} of \cref{Tang18} is affected strongly by their finite-size analysis scheme, in particular in the vicinity of the Gross-Neveu QCP at ${U_c(\gamma)}$, which calls for a revised analysis and interpretation of the numerical data along the lines outlined in this comment.

\begin{acknowledgments}
We thank H.-K. Tang and colleagues for making their data openly available. Funding: This work was supported by FWF project I-2868-N27, FWF project F4018, DFG project RTG 1995 and DFG project FOR 1807.
\end{acknowledgments}

\end{document}